\documentclass{ws-procs9x6}

\usepackage{psfrag}

\begin{document}

\title{Proton-Antiproton Annihilation Into Two Photons at 
Large $s$}

\author{C.~WEISS}

\address{Institut f\"ur Theoretische Physik \\
Universit\"at Regensburg \\ D--93053 Regensburg, Germany \\
E-mail: christian.weiss@physik.uni-regensburg.de}  


\maketitle

\abstracts{Exclusive proton--antiproton annihilation into two photons can 
be viewed as the Compton process in the crossed channel. At 
large $s$ $(\approx 10 \, {\rm GeV}^2 )$ and $|t|, |u| \sim s$ this process 
can be described by a generalized partonic picture, analogous to the 
``soft mechanism'' in wide--angle real Compton scattering. The two photons 
are emitted in the annihilation of a single fast quark and antiquark
(``handbag graph''). The transition of the $p \bar p$ system to a 
$\bar q q$ pair through soft interactions is described by
double distributions, which can be related to the timelike proton elastic 
form factors as well as, by crossing symmetry, to the usual quark--antiquark 
distributions in the nucleon. We estimate that this reaction should be 
observable with reasonable statistics at the proposed 
$1.5\ldots 15\, {\rm GeV}$ high--luminosity antiproton storage ring 
(HESR) at GSI.
(Talk presented at the Workshop on Exclusive Processes at High Momentum 
Transfer, Jefferson Lab, Newport News, VA, May 15--18, 2002)}

Compton scattering, {\it i.e.}, the scattering of real or virtual photons
off hadrons, has proven to be a useful tool for investigating
the structure of the nucleon. Of particular interest are certain 
extreme kinematical situations where
the reaction mechanism simplifies. In the limit of large 
virtuality of the incoming photon and fixed $t$ (``deeply virtual Compton 
scattering'') QCD implies that the amplitude factorizes into a 
hard photon--quark amplitude and a generalized parton distribution,
representing the information about the structure of the nucleon. 
Similarly, it has been argued that real Compton scattering at large 
$s$ $(\approx 10 \, {\rm GeV}^2 )$ and $|t|, |u| \sim s$ (wide--angle 
scattering) is dominated by contributions in which the photon scatters 
off a single quark/antiquark in the nucleon
(``handbag graph'').\cite{Radyushkin:1998rt,Diehl:1998kh,Huang:2001ej} 
The soft interactions responsible for the emission and absorption of 
the active quark/antiquark by the nucleon are parametrized by double 
distributions, which can be related to both the usual quark/antiquark 
distributions measured in inclusive deep--inelastic scattering
and the elastic form factors of the proton. This so-called ``soft mechanism''
describes well the existing data for the total cross section,\cite{Shupe:vg}
and also recent results for the spin asymmetry from the JLAB Hall A 
experiment.\cite{JLAB} The hard scattering 
mechanism, in which the struck quarks rescatters via gluon exchange of
virtuality $\sim t$, is relevant only at asymptotically large $t$ and 
falls short of the measured cross section at JLAB 
energies.\cite{Radyushkin:1998rt}
\par
Exclusive proton--antiproton annihilation into two photons,
$p \bar p \rightarrow \gamma\gamma$, can be regarded as the Compton
process in the crossed channel. This reaction could be studied with the
proposed high--luminosity $1.5\ldots 15 \, {\rm GeV}$ antiproton storage 
ring (HESR) at GSI.\cite{GSI} In this talk I would like to argue that 
at large $s$ and $|t|, |u| \sim s$ (wide--angle scattering)
this process can be described by a generalized partonic picture analogous 
to the ``soft mechanism'' in wide--angle real Compton scattering.
The results reported here have been obtained in collaboration with
A.~Freund (Regensburg U.), A.~V.~Radyushkin (Jefferson Lab and
Old Dominion U.), and A.~Sch\"afer (Regensburg U.). A detailed 
account will be published shortly.\cite{paper} Similar ideas have 
been presented by P.~Kroll at this meeting.\cite{Kroll02,Diehl:2001fv}
%
%
\begin{figure}[t]
\psfrag{mys}{{\Large $s$}}
\psfrag{crossed}{\Large + crossed}
\includegraphics[width=8.5cm,height=2.2cm]{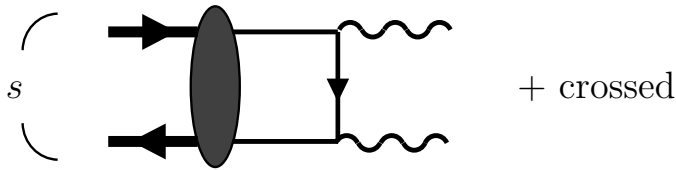}
\caption[]{The ``handbag'' contribution to 
$p \bar p \rightarrow \gamma\gamma$ annihilation}
\label{fig_handbag}
\end{figure}
\par
To motivate the partonic picture of proton--antiproton annihilation
let us recall $e^+e^-$ annihilation into two photons in QED. This process 
proceeds via the $t$ (or $u$) channel exchange of a virtual electron/positron 
with spacelike momentum. In $p \bar p \rightarrow \gamma \gamma$
annihilation in QCD, the exchanged system consists of at least three quarks.
At large momentum transfer such an exchange should be strongly 
suppressed by the proton and antiproton wave functions. In this situation 
the most efficient way of accommodating a large momentum transfer is
the handbag diagram shown in Fig.~\ref{fig_handbag}. In the first part
the proton--antiproton system makes a transition to a quark--antiquark 
pair by exchanging a virtual $qq$ (``diquark'') type system, denoted by 
the blob in the diagram, whose spacelike virtuality is limited by the 
bound--state wave functions. In the second part, the quark--antiquark pair 
annihilates into two photons by exchanging a highly virtual quark/antiquark, 
exactly as in $e^+e^-$ annihilation in QED. This picture is consistent:
Because of the limit on the virtuality of the exchanged diquark--type system 
the active quark (antiquark) carries a significant fraction of the proton 
(antiproton) momentum, which in turn makes for a large virtuality
in the quark propagator connecting the photon vertices. In short, our 
picture states that the dominant contribution
to exclusive $p \bar p \rightarrow \gamma \gamma$ at large $s, |t|$ 
and $u$ comes from the annihilation of ``fast'' quarks and antiquarks 
in the proton and antiproton, respectively.
%
%
\begin{figure}[t]
\psfrag{p+r/2}{{\large $p+r/2$}}
\psfrag{p-r/2}{{\large $p-r/2$}}
\psfrag{k1}{{\large $(1 + \alpha)p + \tilde x r/2$}}
\psfrag{k2}{{\large $(1 - \alpha)p - \tilde x r/2$}}
\includegraphics[width=5.4cm,height=2.2cm]{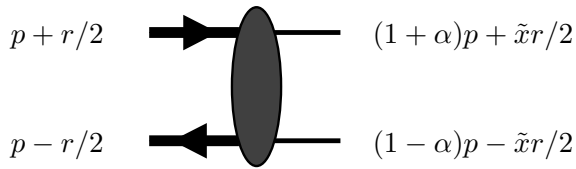}
\caption[]{Graphical representation of the double distribution
describing the transition of the $p\bar p$ system to a $q \bar q$ pair.}
\label{fig_dd}
\end{figure}
\par
The matrix element describing the transition of the $p\bar p$ system 
to a $q \bar q$ pair through ``soft'' interactions can be parametrized 
by double distributions, which are the timelike analogue of the functions 
parametrizing the matrix element in wide--angle Compton 
scattering. These distributions
measure the momenta of the active quark and antiquark in terms of the
proton and antiproton momenta, or, equivalently, their
average and difference, $p$ and $r$, see Fig.~\ref{fig_dd}.
Transverse momenta are neglected here, and the variables $\tilde x$ and
$\alpha$ obey the ``partonic'' restrictions $|\tilde x| + |\alpha| < 1$.
There are two types of matrix elements, corresponding to
Dirac structures $(\gamma_\mu)_{ij}$ and $(\gamma_\mu \gamma_5 )_{ij}$
in the quark spinor indices. Following 
Refs.\cite{Radyushkin:1999es}, we parametrize them by 
two double distributions $F_a (\tilde x, \alpha ; s)$ and 
$G_a (\tilde x, \alpha ; s)$, depending on the partonic variables
$\alpha$ and $\tilde x$ as well as on $s$ (the subscript $a = u, d$ 
represents the quark flavor); details will be given elsewhere.\cite{paper}
When integrating over the spectral variables and summing over the 
contributions of the different quark flavors, the new double distribution
$F_a (\tilde x, \alpha ; s)$ reduces to the timelike proton Dirac form 
factor,
\begin{equation}
\sum_a e_a^2
\int\hspace{-1.5em}\int\limits_{|\tilde x| + |\alpha| < 1} \hspace{-1em} 
d\tilde x \, d\alpha \; F_a (\tilde x, \alpha; s) 
\;\; = \;\;  F_{1} (s) ;
\label{reduction_FF}
\end{equation}
a similar relation holds for $G_a (\tilde x , \alpha ; s)$ and the axial
nucleon form factor. (For simplicity we neglect here components of the 
matrix element corresponding to the Pauli and pseudoscalar form 
factors; these components should be included in a more complete
treatment.) Another useful relation is found in the limit $s \rightarrow 0$, 
where one can use crossing symmetry to relate the annihilation--type 
double distribution of Fig.~\ref{fig_dd} to the usual scattering--type
double distributions of wide--angle Compton scattering.
In particular, this implies
\begin{equation}
\int\limits_{-1 + |\tilde x|}^{1 - |\tilde x|}
\!\!\! d\alpha \; 
F_a (\tilde x, \alpha; s = 0)
\;\; = \;\;  f_a (\tilde x ) ,
\label{reduction_PDF}
\end{equation}
where $f_a (\tilde x ) = \theta(\tilde x) q_a (\tilde x) - 
\theta(-\tilde x) \bar q_a (-\tilde x)$ is the usual unpolarized
quark/antiquark distribution of flavor $a$
in the proton, as measured in deep--inelastic scattering. 
In a similar way the function $G_a (\tilde x , \alpha ; s)$ reduces to the 
polarized distribution. These ``reduction relations'' provide
constraints for models of the double distributions.
\par
To construct an explicit model for the double distributions we 
factorize them into an $s$--independent double distribution
and a ``cutoff function'' containing the $s$--dependence,
\begin{equation}
F_a (\tilde x, \alpha; s) \;\; = \;\; f_a (\tilde x, \alpha )
\; S (\tilde x, \alpha; s) .
\label{factorized_ansatz}
\end{equation}
The $s$--independent double distribution we model as 
$f_a (\tilde x, \alpha ) = f_a (\tilde x) h (\tilde x, \alpha )$,
where $f_a (\tilde x)$ is the usual quark/antiquark distribution,
and $h(\tilde x, \alpha )$ a normalized profile 
function; a particularly simple choice is 
$h(\tilde x, \alpha ) = \delta (\alpha )$.\cite{Radyushkin:1999es} 
The distributions
refer to a scale of the order $|t| \sim 1 \,{\rm GeV^2}$. The cutoff function 
$S (\tilde x, \alpha; s)$, which is defined to be unity at $s = 0$,
implements the restriction on the virtuality of the exchanged 
``diquark''--type system in the handbag diagram of Fig.~\ref{fig_handbag}:
\begin{equation}
\frac{[(1 - \tilde x)^2 - \alpha^2] s}
{4 \tilde x (1 - \tilde x )}
\; < \;  \lambda^2 ,
\label{S_new}
\end{equation}
where $\lambda^2$ is a parameter. For the scattering--type double 
distributions appearing in wide--angle Compton scattering this
cutoff could be derived from the overlap of light--cone wave 
functions;\cite{Radyushkin:1998rt,Diehl:1998kh} such an interpretation 
is no longer possible in the annihilation channel. In practice, 
we choose a Gaussian cutoff and fix the parameter $\lambda^2$ by 
fitting the proton form factor, {\it cf.}\ 
Eq.(\ref{reduction_FF}); for details we refer to the 
forthcoming publication.\cite{paper}
\par
\begin{figure}[t]
\psfrag{RAV2}{{\Large $R_{V,A}^2$}}
\psfrag{s/GeV2}{{\Large $s / {\rm GeV}^2$}}
\includegraphics[width=8.4cm,height=5.95cm]{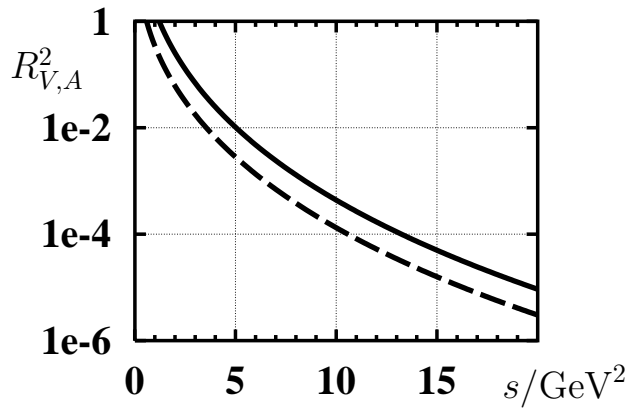}
\caption[]{The form factors $R_V^2 (s)$ (solid line)
and $R_A^2 (s)$ (dashed line) calculated from the double distribution 
model, {\it cf.}\ Eq.(\ref{R_def}).}
\label{fig_sdep}
\end{figure}

The helicity--averaged differential cross section obtained from 
our simple model is 
\begin{equation}
\frac{d\sigma}{d\cos\theta} 
\;\; = \;\; 
\frac{2\pi \alpha^2_{\rm em}}{s} \; 
\frac{R_V^2 (s) \cos^2 \theta + R_A^2 (s)}{\sin^2 \theta} ,
\label{cross}
\end{equation}
where $\theta$ is the scattering angle in the center--of--mass system.
The information about the structure of the proton is contained in 
generalized form factors, defined as
\begin{equation}
R_V (s) \;\; \equiv \;\; \sum_a e_a^2
\int\hspace{-1.5em}\int\limits_{|\tilde x| + |\alpha| < 1} \hspace{-1em} 
d\tilde x \, d\alpha \; 
\frac{F_a (\tilde x, \alpha ; s)}{\tilde x} ;
\label{R_def}
\end{equation}
$R_A (s)$ is given by the corresponding integral over the double distribution
$G_f (\tilde x, \alpha ; s)$. The integral here differs from the
one in the proton elastic form factor, Eq.(\ref{reduction_FF}), 
by an additional factor $1/\tilde x$ in the integrand; this is just the
``remnant'' of the quark propagator in the hard 
$q\bar q \rightarrow \gamma\gamma$ scattering amplitude in
the handbag graph of Fig.~\ref{fig_handbag}. For 
$R_V(s) \equiv R_A(s) \equiv 1$ 
the expression (\ref{cross}) would reproduce the Klein--Nishina formula
for the $e^+e^-\rightarrow \gamma\gamma$ cross section in QED.
Note that our partonic picture is applicable only for 
$|t|, |u| \sim s$, which implies that $\theta$ should be sufficiently far 
from $0$ or $\pi$
(wide--angle scattering). The numerical results for the form factors 
are shown in Fig.~\ref{fig_sdep}.
%
%
\par
It is interesting to make a quick estimate of the counting rate 
for $p\bar p \rightarrow \gamma\gamma$ annihilation expected for
the proposed $1.5\ldots 15 \, {\rm GeV}$ antiproton storage ring (HESR) 
at GSI.\cite{GSI} With a fixed solid target the luminosity could be 
as high as $L = 2 \times 10^{32} \, {\rm cm}^{-2} \, {\rm s}^{-1}
= 2 \times 10^6 \, {\rm fm}^{-2} \, s^{-1}$. Since our partonic picture 
applies only for $|t|, |u| \sim s$ we have to exclude the region 
of $\theta$ close to $0$ or $\pi$. For 
$s = 10\,{\rm GeV^2}$, the range $45^\circ < \theta < 135^\circ$ 
would correspond to $|t|, |u| > 1.5\, {\rm GeV^2}$, which is
hopefully sufficient for our approximations to work. The integrated 
cross section over this region is $1.3 \times 10^{-10}\, {\rm fm}^2$, 
which corresponds to a counting rate of $2.6 \times 10^{-4} {\rm sec}^{-1}$, 
that is $\sim 700$ events per month. Note that already slightly lower 
values of $s$ would increase the counting rate dramatically, {\it cf.}\
Fig.~\ref{fig_sdep}. The process should thus be measurable 
with reasonable statistics at the proposed facility.
\par
To summarize, we have discussed exclusive annihilation 
$p\bar p \rightarrow \gamma\gamma$ in a generalized partonic picture.
Our approach represents an attempt to extend the ``soft mechanism'',
which successfully describes wide--angle Compton scattering at JLAB
energies, to the annihilation channel. The timelike double distributions
describing the transition of the $p\bar p$ system to a $q \bar q$ pair
are strongly constrained by data for the timelike proton form factor and
the quark/antiquark distributions in the proton, which have been 
obtained from independent measurements. Our estimate of the cross
section, based on a simple model for the double distributions, 
suggests that this reaction could be observed with reasonable statistics
at the proposed GSI HESR facility. This would offer the exciting 
possibility of studying the hadronic Compton process in the annihilation
channel. The results could also be compared to data on hadron production 
in photon--photon collisions observed in $e^+e^-$ 
experiments.
\par
C.W.\ is supported by a Heisenberg Fellowship from 
Deutsche Forschungsgemeinschaft (DFG).

\begin{thebibliography}{0}
%
%
\bibitem{Radyushkin:1998rt}
A.~V.~Radyushkin,
Phys.\ Rev.\ D {\bf 58}, 114008 (1998).
%
%
\bibitem{Diehl:1998kh}
M.~Diehl, T.~Feldmann, R.~Jakob and P.~Kroll,
Eur.\ Phys.\ J.\ C {\bf 8}, 409 (1999).
%
%
\bibitem{Huang:2001ej}
H.~W.~Huang, P.~Kroll and T.~Morii,
Eur.\ Phys.\ J.\ C {\bf 23}, 301 (2002).
%
%
\bibitem{Shupe:vg}
M.~A.~Shupe {\it et al.},
Phys.\ Rev.\ D {\bf 19}, 1921 (1979).
%
%
\bibitem{JLAB} A.~Nathan, Talk presented at the Workshop on Exclusive 
Processes at High Momentum Transfer, Jefferson Lab, Newport News, 
VA, May 15--18, 2002
%
%
\bibitem{GSI} ``An International Accelerator Facility for Beams
of Ions and Antiprotons'', GSI Conceptual Design Report, 
November 2001
%
%
\bibitem{paper} A.~Freund, A.~V.~Radyushkin, A.~Sch\"afer,
and C.~Weiss, to be published.
%
%
\bibitem{Kroll02}  P.~Kroll, Talk presented at the Workshop on Exclusive 
Processes at High Momentum Transfer, Jefferson Lab, Newport News, 
VA, May 15--18, 2002
%
%
\bibitem{Diehl:2001fv}
See also: M.~Diehl, P.~Kroll and C.~Vogt,
Phys.\ Lett.\ B {\bf 532}, 99 (2002).
%
%
\bibitem{Radyushkin:1999es}
A.~V.~Radyushkin,
Phys.\ Rev.\  {\bf D59} (1999) 014030;
Phys.\ Lett.\  {\bf B449} (1999) 81.
%
%
\end{thebibliography}
\end{document}